\documentclass[reqno]{amsart}
\usepackage{url}
\usepackage{amssymb,amsthm,amsfonts,amstext}
\usepackage{amsmath}
\usepackage{mathptmx}       
\usepackage{helvet}         
\usepackage{courier}        
\usepackage{graphicx}
\usepackage{enumerate}
\usepackage[active]{srcltx}
\usepackage[brazil,english]{babel}
\usepackage[latin1]{inputenc}
\usepackage{color}

\newtheorem{theorem}{Theorem}

\newtheorem{lemma}{Lemma}

\let\k=\kappa

\title{ Generic properties for random repeated quantum iterations}

\author{ Artur O. Lopes and Marcos Sebastiani}

\begin{document}

\maketitle

\begin{abstract}

We denote by $M^n$ the set of $n$ by $n$ complex matrices.
Given a fixed density matrix $\beta:\mathbb{C}^n \to \mathbb{C}^n$ and a fixed unitary operator $U :  \mathbb{C}^n \otimes \mathbb{C}^n \to
 \mathbb{C}^n \otimes \mathbb{C}^n$, the transformation $\Phi: M^n \to M^n$
 $$ Q \to  \Phi (Q) =\, \text{Tr}_2 (\,U \,  ( Q \otimes \beta )\, U^*\,)$$
 describes the interaction of $Q$ with the external source $\beta$. The result of this is $\Phi(Q)$. If $Q$ is a density operator then $\Phi(Q)$ is also
 a density operator. The main interest is to know what happen when we repeat several times the action  of
 $\Phi$ in an initial fixed density operator $Q_0$. This procedure is known as random repeated quantum iterations and is of course related to the existence of one or more fixed points for $\Phi$.

 In \cite{NP}, among  other things,  the authors show that for a fixed $\beta$ there exists a set of full probability for the Haar measure such that the unitary operator $U$ satisfies the property that for the associated $\Phi$ there is a unique fixed point $ Q_\Phi$. Moreover, there exists convergence of the iterates $\Phi^n (Q_0) \to Q_\Phi$, when $n \to \infty$,  for any given $Q_0$

We show here  that there is an open and dense set of unitary operators $U: \mathbb{C}^n  \otimes \mathbb{C}^n \to  \mathbb{C}^n  \otimes \mathbb{C}^n $ such that the associated $\Phi$ has a unique fixed point.

 We will also consider a detailed analysis of the case when $n=2$. We will be able to show explicit results. We consider the $C^0$ topology on the coefficients of $U$. In this case we will exhibit the explicit expression on the coefficients of  $U$ which assures the existence of a unique fixed point for $\Phi$. Moreover, we present the explicit expression of the fixed point $Q_\Phi$

\end{abstract}

\section{Introduction}

We denote by $M^n$ the set of $n$ by $n$ complex matrices.
Given a fixed density matrix $\beta:\mathbb{C}^n \to \mathbb{C}^n$ and a fixed unitary operator $U :  \mathbb{C}^n \otimes \mathbb{C}^n \to
 \mathbb{C}^n \otimes \mathbb{C}^n$, the transformation $\Phi: M^n \to M^n$
 $$ Q \to  \Phi (Q) =\, \text{Tr}_2 (\,U \,  ( Q \otimes \beta )\, U^*\,)$$
 describes the interaction of $Q$ with the external source $\beta$.

We assume that all eigenvalues of $\beta $ are strictly positive.

 In \cite{NP} the model is precisely explained:  $Q$ is in the small system and $\beta$ describes the environment. Then
$ \Phi (Q)$ gives the output of the action of $\beta$ in $Q$ given the action of the unitary operator $U$.

Other related papers are \cite{Liu} and \cite{BJK}.

The main question is about the convergence of the iterates $\Phi^n (Q_0) $, when $n \to \infty$,  for any given $Q_0$. It is natural to expect that any limit (if exists) is a fixed point for $\Phi$.

\medskip

Our purpose is to show the following theorem:
\begin{theorem}
Given a fixed density matrix $\beta:\mathbb{C}^n \to \mathbb{C}^n$, for an open and dense set of  unitary operators $U :  \mathbb{C}^n \otimes \mathbb{C}^n \to
 \mathbb{C}^n\otimes \mathbb{C}^n$ the transformation $\Phi: M^n \to M^n$
 $$ Q \to  \Phi (Q) =\, \text{Tr}_2 (\,U \,  ( Q \otimes \beta )\, U^*\,)$$
has a unique fixed point $Q_\Phi$. In the case $n=2$ we present explicitly  the analytic characterization of such family of $U$ and also the explicit formula for $Q_\Phi$.
\end{theorem}

\bigskip

This result implies one  of the main results in \cite{NP} that we mentioned before.

\section{The general dimensional case}

Suppose $V$ is a complex Hilbert space of dimension n $\geq 2$ and $\mathcal{L}(V)$ denotes the space of linear transformations of $V$ in itself.

Then, $Tr_2:\mathcal{L}(V\otimes V)\to \mathcal{L}(V)$, given by
 $Tr_2 (A \otimes B)= Tr (B)\, A$. 
 
 There is  a canonical way to extend the inner product on $V$ to $V \otimes V$.

 We fix a density matrix $\beta \in \mathcal{L}(V)$. For each unitary operator $U \in \mathcal{L}(V\otimes V)$ we denote by $\Phi_U:  \mathcal{L}(V)\to  \mathcal{L}(V) $ the linear transformation
$$\Phi_U(A) = Tr_2 \,( U \, ( \otimes \beta)\, U^*).$$

We denote by $\Gamma\subset  \mathcal{L}(V)$ the set of density operators. It will be shown  that $\Phi_U$ preserves $\Gamma$. As $\Gamma$ is a convex compact space it has a fixed point.

The set of unitary operators is denoted by $\mathcal{U}$.

If $A$ is such that $\Phi_U (A)=A$, then it follows that the range of $ \phi_U - I$ is smaller or equal  to $n^2-1$.

We will show that there exist a proper real analytic subset $X \subset \mathcal{U} $   such that if $U$ is not in $X$, then range $\Phi_U - I=n^2 -1$. In this case the fixed point is unique. More precisely
$$ X =\{ U \in \mathcal{U} \,:\, \text{range}\, (\Phi_U - I)< n^2 -1\}.$$

This $X\subset \mathcal{U}$ is an analytic set because is described  by equations given by the determinant of minors equal to zero. It is known that the complement of an analytic set, also known as a Zariski open set,  is empty or is open and dense on the analytic manifold (see \cite{Real}). Therefore, in order to prove our main result we have to present an explicit $U$ such that range of $\, (\Phi_U - I)$ is $ n^2 -1.$

This will be the purpose of our reasoning described below.

The bilinear transformation $(A,B) \to Tr (B) A$ from  $\mathcal{L}(V)\times \mathcal{L}(V)$  to $\mathcal{L}(V)$ induces the linear transformation
$$ Tr_2 : \mathcal{L}(V\otimes V) = [\mathcal{L}(V) \otimes \mathcal{L}(V)]\,\to \mathcal{L}(V)      .    $$

Denote by $e_1,e_2,...,e_n$ an orthonormal basis for $V$. We also denote $L_{ij}\in \mathcal{L}(V)$ the transformation such that $L_{ij}(e_j)=e_i$ and $L_{ij}(e_k)=0$ if $k \neq j$.

The $L_{ij}$ provides a basis for $\mathcal{L}(V)$.

If $A \in \mathcal{L}(V)$ we can write $A= \sum_{i,j} \, a_{ij} \, L_{ij}$ and we call $[a_{ij}]_{1\leq i,j\leq n}$ the matrix of $A$.

Note that $e_i\otimes e_j$, $1\leq i,j\leq n$ is an orthonormal  basis of $V \otimes V$. Moreover,
$$ L_{ik} \otimes L_{jl} (e_k\otimes e_l )= e_i\otimes e_j,$$
and
$$ L_{ik} \otimes L_{jl} (e_p\otimes e_q )= 0\,\,\text{if}\,\, (p,q)\neq (k,l).$$

It is also true that:

a) $L_{ij}\, L_{pq}=0$ if $j \neq p$,

b) $L_{ij}\, L_{pj}=L_{iq},$

c) Tr $(L_{ij})=0$ if $i\neq j$ and  Tr $(L_{ii})=1$.

One can see that $L_{ik}\,\otimes L_{jl}$, $1\leq i,k,j,l\leq n$ is a basis for $\mathcal{L}(V\otimes V).$

Given $T\in  \mathcal{L}(V\otimes V)$ denote $T= \sum t_{i,j,k,l} L_{ik}\,\otimes L_{jl}$. Then,
$$ Tr_2 (T) = \sum t_{i,j,k,j} L_{ik}=  \sum_{ik} \,(\sum_j t_{i,j,k,j} \,)\,L_{ik}.$$

In the appendix we give a direct proof that:  if $A \in \Gamma$, then $\Phi_U(A)\in \Gamma$, for all $U \in \mathcal{U}.$

\bigskip

Now we will express $\Phi_U$ in coordinates. We choose  an orthonormal base $e_1,e_2,..,e_n\in V$ which diagonalize $\beta$. That is
$$\beta= \sum_q\, \lambda_q L_{qq},\,\,\,\lambda_q >0,\,\,\,1\leq q\leq n,\,\,\sum_q \lambda_q=1.$$

Given $r,s$, $1\leq r,s\leq n$, we will calculate $\Phi_U (L_{rs})$.

Suppose $U= \sum u_{i,j,k,l} L_{ik}\,\otimes L_{jl}$, then $U^*= \sum \overline{u_{i,j,k,l}}\, L_{ik}\,\otimes L_{jl}$ and
$$ (L_{rs}\otimes \beta)\, U^* = (\sum_q \lambda_q\, L_{rs} \otimes L_{qq} )\, U^*=\sum_j \lambda_j \,\,\overline{u_{k,l,s,j}}\,\,\, L_{rk} \otimes L_{jl} .$$

Now, we write
$U= \sum u_{\alpha,\beta,\gamma,\delta} L_{\alpha\, \gamma}\,\otimes L_{\beta\, \delta}$. Then, we get

$$U (L_{rs}\otimes \beta)\, U^*\,=\, \sum \lambda_j \,\, u_{\alpha,\beta,r,j}\,\, \overline{u_{k,l,s,j}}\,L_{\alpha\, k} \otimes L_{\beta\,l}.$$

Finally,
$$ \Phi_U (L_{rs})\,=\, \sum \lambda_j \,\, u_{\alpha,l,r,j}\,\, \overline{u_{k,l,s,j}}\,\,L_{ \alpha k}=  \sum_{\alpha,k}\,\,(\, \sum_{j,l}  \lambda_j \,\, u_{\alpha,l,r,j}\,\, \overline{u_{k,l,s,j}}\,\,\lambda_j\,\,)\,\,L_{ \alpha k}.$$

As $\Gamma$ is convex and compact and $\phi_U$ is continuous as we said before   there exist a fixed point $A \in \Gamma$. In particular the range of $\phi_U$ is smaller or equal to $n^2-1.$

We will present an explicit $U$ such that range of $\, (\Phi_U - I)$ is $ n^2 -1.$

This will be described by  a certain kind of circulant unitary operator

 \medskip

Suppose $u_1,u_2,...,u_{n^2}$ are complex number of modulus $1$. We define $U$ in the following way
$$ U(e_1\otimes e_1)= u_1\, (e_1 \otimes e_2),\,\, U(e_1\otimes e_2)= u_2\, (e_1 \otimes e_3),..., U(e_1\otimes e_n)=u_n\, (e_2\otimes e_1),$$

$$ U(e_2\otimes e_1)= u_{n+1}\, (e_2 \otimes e_2),\,\, U(e_2\otimes e_2)= u_{n+2}\, (e_2 \otimes e_3),..., U(e_2\otimes e_n)=u_{2n}\, (e_3\otimes e_1),$$

$$....$$

$$ U(e_n\otimes e_1)= u_{n^2 -n+1}\, (e_n \otimes e_2),\,\, U(e_n\otimes e_2)= u_{n^2-n +2}\, (e_n \otimes e_3),..., U(e_n\otimes e_n)=u_{n^2}\, (e_1\otimes e_1),$$

We will show that for some convenient choice of  $u_1,u_2,...,u_{n^2}$ we will get that  the range of $\Phi_U-I$ is $n^2 -1$.

Suppose

$$U= \sum u_{i,j,k,l}\,\, L_{ik}\,\otimes L_{jl}   , $$
in this case
$$U(e_k\otimes e_l)= \sum_{i,j}\,\, u_{i,j,k,l}\,\, e_{i}\,\otimes\, e_{j}   . $$

By definition of $U$ we get

a) if $l<n$, then  $ u_{i,j,k,l}\neq 0$, if and only if,  $i=k$, $j=l+1$;

b) if $k<n$, then  $ u_{i,j,k,n}\neq 0$, if and only if,  $i=k+1$, $j=1$;

c) $ u_{i,j,n,n}\neq 0$, if and only if,  $i=j=1$.

\medskip

For fixed $r,s$ such that  $1\leq ,r,s\leq n$ we get from a), b) and c):

$1\leq r <n, $ $1\leq s<n$, implies
$$\Phi_U(L_{rs})=(\,\sum_{j=1}^{n-1}\, u_{r,j+1,r,j\, }\,\,\, \overline{u_{s,j+1,s,j}}\,\,\lambda_j\, \,)\,L_{rs}\,+\,u_{r+1,1,r,n\, }\,\,\, \overline{u_{s+1,1,s,n}}\,\,\lambda_n\, \,\,L_{(r+1)(s+1)}, $$

\medskip

$1\leq s<n$, implies
$$\Phi_U(L_{ns})=(\,\sum_{j=1}^{n-1}\, u_{n,j+1,n,j\, }\,\,\, \overline{u_{s,j+1,s,j}}\,\,\lambda_j\, \,)\,L_{ns}\,+\,u_{1,1,n,n\, }\,\,\, \overline{u_{s+1,1,s,n}}\,\,\lambda_n\, \,\,L_{1\,(s+1)}, $$

\medskip

$1\leq r<n$, implies
$$\Phi_U(L_{rn})=(\,\sum_{j=1}^{n-1}\, u_{r,j+1,r,j\, }\,\,\, \overline{u_{n,j+1,n,j}}\,\,\lambda_j\, \,)\,L_{rn}\,+\,u_{r+1,1,r,n\, }\,\,\, \overline{u_{1,1,n,n}}\,\,\lambda_n\, \,\,L_{(r+1)\,1}. $$

In particular for $1 \leq r<n$ we have $\Phi_U(L_{rr})=(1-\lambda_n)\, L_{rr}+ \lambda_n L_{(r+1)\,(r+1)}.$

In order to show that the range of $\Phi_U-I$ is $n^2-1$ we will show that the $\phi_U(L_{rs})-L_{rs}$ are linearly independent for $(r,s)\neq (n,n)$

Suppose that
$$ \sum_{(r,s)\neq (n,n)} c_{rs}\, (\phi_U(L_{rs})-L_{rs})=0. $$

The coefficient of $L_{11} $ is $-\lambda_n \, c_{11}$, then $c_{1\,1}=0.$

The coefficient of $L_{22} $ is $\lambda_n \, c_{11}- \lambda_n c_{22}$, then $c_{2\,2}=0.$

$...$

The coefficient of $L_{nn} $ is $\lambda_n \, c_{(n-1)\,(n-1)}$, then $c_{(n-1)\,(n-1)}=0.$

Then, we get that
\begin{equation} \label{oo}
\sum_{r\neq s} c_{rs}\, (\phi_U(L_{rs})-L_{rs})=0. 
\end{equation}

We will divide the proof in several different cases.

a) Case $n=2$.

$$ \sum_{r\neq s} c_{rs}\, (\phi_U(L_{rs})-L_{rs})=c_{12}\, (\phi_U(L_{12})-L_{12}) + c_{21}\, (\phi_U(L_{21})-L_{21}). $$

By definition of $U$ we have that $   u_{1,2,1,1}=u_1, $ $   u_{2,1,1,2}=u_2, $ $   u_{2,2,2,1}=u_3, $ $   u_{1,1,2,2}=u_4. $

Therefore,

$$\phi_U (L_{12})- L_{12}\,=\, (u_1\, \overline{u_3}\, \lambda_1 \, -1) L_{12} +  u_2\, \overline{u_4}\, \lambda_2 \,  L_{21}$$
and

$$\phi_U (L_{21})- L_{21}\,=\, (u_3\, \overline{u_1}\, \lambda_1 \, -1) L_{21} +  u_4\, \overline{u_2}\, \lambda_2 \,  L_{12}.$$

From (\ref{oo}) it follows that

$$ (u_1\, \overline{u_3}\, \lambda_1 -1)\, c_{12} + u_4 \overline{u_2}\, \lambda_2 c_{21}=0$$

$$ u_2 \overline{u_4}\, \lambda_2 c_{12} + (u_3\, \overline{u_1}\, \lambda_1 -1)\, c_{21} =0.$$

Taking $U$ such that $u_1=i$, $u_2=u_3=u_4=1$ it is easy to see that the determinant of the above system is not equal to zero. Then we get that $c_{12}=c_{21}=0.$

Then, we get a $U$ with maximal range.

\medskip

b) Case $n>2$.

We choose  $u_1,u_2,...,u_{n^2}$ according to Lemma \ref{pp} below.

The equations we consider before can be written as

$1\leq r<n$, $1\leq s<n$,  $r\neq s$, then, $\Phi_U(L_{r\,s}) -L_{r\,s} = (a_{r\,s} -1) \,L_{r\,s} + b_{r\,s} \,L_{(r+1)\, (s+1)}, $

$1\leq s<n$, then, $\Phi_U(L_{n\,s}) -L_{n\,s} = (a_{n\,s} -1) \,L_{n\,s} + b_{n\,s} \,L_{1\, (s+1)}, $

$1\leq r<n$, then, $\Phi_U(L_{r\,n}) -L_{r\,n} = (a_{r\,n} -1) \,L_{r\,n} + b_{r\,n} \,L_{(r+1)\, 1}.$

For instance 

$$a_{rs} = \sum_{j=1}^{n-1}\, u_{r,j+1,r,j\, }\,\,\, \overline{u_{s,j+1,s,j}}\,\,\lambda_j\, \,,$$

and 
$$b_{rs} =u_{r+1,1,r,n\, }\,\,\, \overline{u_{s+1,1,s,n}} \, \lambda_n.$$

Note that $u_{r,j+1,r,j\, }\,\,\, \overline{u_{s,j+1,s,j}}$ has modulus one and also $ u_{r+1,1,r,n\, }\,\,\, \overline{u_{s+1,1,s,n}}$.

Moreover, $|\,b_{r\,s}\, |= \lambda_n>0$ and $|\,a_{r\,s}\,|< \lambda_1 +...+\lambda_{n-1}$.  Indeed, note first that the products  $u_{r,j+1,r,j\, }\,\,\, \overline{u_{s,j+1,s,j}}$ are different by the choice of the $u_{i,j,k,l}$ (see Lemma \ref{pp}).  Furthermore,  by Lemma \ref{pp2} we get that $|a_{rs} |$ can not be equal to $\lambda_1 +...+\lambda_{n-1}$.

Therefore, $|\, a_{r\,s}-1\,| \geq 1 - |a_{r\,s}|>1 - \sum_{q=1}^{n-1} \lambda_q=\lambda_n =|\, b_{i\,j}\,|>0,$ for all $r,s,i,j$ and $r\neq s$, $i\neq j.$

\medskip

Suppose $2\leq k\leq n.$

Remember that the $L_{ij}$ define a linear independent set.

The coefficient of $L_{1\,k}$ in (\ref{oo}) is 
$$c_{1\,k}\,(a_{1\,k}-1)\,+\, c_{n\, (k-1) }\, b_{n\, (k-1)}=0.$$

The coefficient of $L_{n\,(k-1)}$ in (\ref{oo}) is 
$$c_{n\,(k-1)}\,(a_{n\,(k-1)}-1)\,+\, c_{(n-1)\, (k-2) }\, b_{(n-1)\, (k-1)}=0.$$

The coefficient of $L_{(n-k+2)\,1}$ in (\ref{oo}) is 
$$c_{(n-k+2)\,1}(a_{(n-k+2)\,1}-1)+ c_{(n-k-1)n } b_{(n-k+1) n}=0.$$

The coefficient of $L_{(n-k +1)n}$ in (\ref{oo}) is  
$$c_{(n-k+1)n}(a_{(n-k+1)n}-1)+c_{(n-k)(n-1) }b_{(n-k) (n-1)}=0.$$

The coefficient of $L_{(n-k)\,(n-1)}$ in (\ref{oo}) is 
$$c_{(n-k)\,(n-1)}\,(a_{(n-k)\,(n-1)}-1)\,+\, c_{(n-k-1)\, (n-2) }\, b_{(n-k-1)\, (n-2)}=0.$$

$$...$$

The coefficient of $L_{2\,(k+1)}$ in (\ref{oo}) is 
$$c_{2\,(k+1)}\,(a_{2\,(k+1)}-1)\,+\, c_{1\, k }\, b_{1\, k}=0.$$

If $c_{1\,k}\neq 0$, then, from above we get $|c_{1\,k}|<  |c_{n\,(k-1)}|<...<|c_{2\,(k+1)}|<|c_{1\,k}|.$

Then, we get a contradiction. It follows that $c_{1\,k}=0$.

Therefore,
$$  c_{n\,(k-1)}=  c_{(n-1)\,(k-2)}=... =c_{(n-k+2)\,1}= c_{(n-k+1)\,n}= c_{(n-k)\,(n-1)}=... =c_{2\,(k+1)}=0.$$ 

\bigskip

From this follows that $c_{r\,s}=0$ for all $r,s$, when $r \neq s$. This shows that for such $U$ we have maximal range equal to $n^2-1$.
\bigskip

Now we will prove two Lemmas that we used before.

\begin{lemma} \label{pp}

Given $m \geq 2$, there exists complex numbers $u_1,...,u_m$ of modulus $1$, such that, if $1\leq i \neq j\leq m, $
$1 \leq k \neq l\leq m$ and $u_i\, \overline{u_j} = u_k\, \overline{u_l}$, then $i=k, j=l$.

\end{lemma}

{\bf Proof:} The proof is by induction on $m$

For $m=2$, just take $u_1\, \overline{u_2} $ not in $\mathbb{R}.$

Suppose the claim is true for $m \geq 2$ and $u_1,...,u_m$ the corresponding ones.

Consider 
$$S =\{u_i\,\overline{u_j}\,\,\,|\,\,\, 1 \leq i,j\leq m\   \} $$
and
$$T =\{u_p\,u_q\,\,\,|\,\,\, 1 \leq p,q\leq m\   \}. $$

Then, take $u_{m+1}$ such that $u_{m+1} \overline{u_p} $ is not in $S$ for all $1\leq p\leq m$, and $ u^2_{m+1} $ is not in $T$.

Then, $u_1,...,u_m,u_{m+1}$ satisfies the claim.

\qed

\begin{lemma} \label{pp2} Consider $\lambda_1,...,\lambda_m$, real positive numbers and $z_1,...,z_m$, complex numbers of modulus $1$.

Suppose $|\, \sum_{j=1}^m \, \lambda_j\, z_j\,| = \sum_{j=1}^m \, \lambda_j$, then $z_1=z_2=...=z_m.$

\end{lemma}

{\bf Proof:} The proof is by induction on $m$.

It is obviously true for $m=1$.

Suppose the claim is true for $m-1$ and we will show is true for $m$.

Note that
$$ \sum_{j=1}^m \lambda_j = |\, \sum_{j=1}^m \, \lambda_j\, z_j\,| \leq |\, \sum_{j=1}^{m-1} \, \lambda_j\, z_j\,|  + \lambda_m\leq \sum_{j=1}^m \lambda_j.$$

From, this follows that
$$ |\, \sum_{j=1}^{m-1} \, \lambda_j\, z_j\,|=  \sum_{j=1}^{m-1} \lambda_j .$$

Then, $z_1=z_2=...=z_{m-1}=z$.

Therefore,
$$ \sum_{j=1}^m \lambda_j = |\, z \sum_{j=1}^{m-1} \lambda_j\, +\, z_m\, \lambda_m\,| \leq |\, z \sum_{j=1}^{m-1} \lambda_j\, |\,+\, |z_m\, \lambda_m\,|=  \sum_{j=1}^m \lambda_j.$$

Given $v_1,v_2$ complex numbers such that $|v_1+ v_2|=|v_1| + |v_2|$, then they have the same argument.

Then, there exists an $s>0$ such that $z \, \sum_{j=1}^{m-1} \lambda_j\,=\,s\, z_m \lambda_m$.

Now, taking modulus in both sides of the expression above, we get 
$$\sum_{j=1}^{m-1} \lambda_j=|\,z \, \sum_{j=1}^{m-1} \lambda_j\,|\,=|\,s\, z_m \lambda_m|= s \lambda_m.$$

From this follows that $z_m=z$

\qed

\bigskip

\section{The two dimensional case - explicit results}

Our main interest in this section is  the explicit expression of the $U$ such that the fixed point is unique. We restrict ourselves to the two dimensional case.
\medskip

We will consider a two by two density matrix $\beta$ such that is diagonal in the basis $f_1\in \mathbb{C}^2$, $f_2\in\mathbb{C}^2$.
Without lost of generality we can consider that
$$ \beta =  \left(
\begin{array}{cc}
p_1 & 0\\
0 & p_2
\end{array}
\right),$$

$p_1,p_2>0$.

We will describe initially in coordinates some of the definitions which were used before on the paper.

If

$$R=
\left(
\begin{array}{cc}
R_{11} & R_{12}\\
R_{21} & R_{22}
\end{array}
\right),$$

and
$$S=
\left(
\begin{array}{cc}
S_{11} & S_{12}\\
S_{21} & S_{22}
\end{array}
\right),$$

then

$$R \otimes S=  \left(
\begin{array}{cccc}
R_{11} S_{11} & R_{11}  S_{12}  & R_{12}S_{11} & R_{12}S_{12}\\
R_{11}  S_{21} & R_{11} S_{22}  & R_{12} S_{21}& R_{12}S_{22}\\
R_{21} S_{11} & R_{21}S_{12} & R_{22} S_{11} & R_{22}S_{12}\\
R_{21} S_{21}& R_{21} S_{22}& R_{22}S_{21} & R_{22} S_{22}
\end{array}
\right)  $$

and
$$ \text{Tr}_2 ( R \otimes S)=\left(
\begin{array}{cc}
R_{11} \,(S_{11} + S_{22})  & R_{12}\,(S_{11} + S_{22}) \\
R_{21} \,(S_{11} + S_{22})  & R_{22} \,(S_{11} + S_{22})
\end{array}
\right) .$$

Given
$$T=   \left(
\begin{array}{cccc}
T_{11} & T_{12} &  T_{13} & T_{14}\\
T_{21} & T_{22} &  T_{23} & T_{24}\\
T_{31} & T_{32} &  T_{33} & T_{34}\\
T_{41} & T_{42} &  T_{43} & T_{44}
\end{array}
\right) $$
then, in a consistent way we have
$$ \text{Tr}_2 (T)=  \left(
\begin{array}{cc}
T_{11}+ T_{22} & T_{13}+ T_{24}\\
T_{31}+ T_{42} & T_{33}+ T_{44}
\end{array}
\right)$$

The action of an operator $U$ on $M_2\otimes M_2$ in the basis $e_1 \otimes f_1$, $e_2 \otimes f_1$, $e_1 \otimes f_2$, $e_2 \otimes f_2$
is given by a $4$ by $4$ matrix $U$ denoted by
$$U=   \left(
\begin{array}{cccc}
U_{11}^{11} & U_{11}^{\,12} &  U_{\,12}^{11} & U_{\,12}^{\,12}\\
U_{11}^{\,21} & U_{11}^{22} &  U_{\,12}^{\,21} & U_{\,12}^{22}\\
U_{\,21}^{11} & U_{\,21}^{\,12} &  U_{22}^{11} & U_{22}^{\,12}\\
U_{\,21}^{\,21} & U_{\,21}^{22} &  U_{22}^{\,21} & U_{22}^{22}
\end{array}
\right) $$
and
$$ U^*=   \left(
\begin{array}{cccc}
\overline{U_{11}^{11}} & \overline{U_{11}^{\,21}} & \overline{ U_{\,21}^{11}} & \overline{U_{\,21}^{\,21}}\\
\overline{U_{11}^{\,12}} & \overline{U_{11}^{22}} &  \overline{U_{\,21}^{\,12}} &\overline{ U_{\,21}^{22}}\\
\overline{U_{\,12}^{11}} & \overline{U_{\,12}^{\,21}} & \overline{ U_{22}^{11}} & \overline{U_{22}^{\,21}}\\
\overline{U_{\,12}^{\,12}} & \overline{U_{\,12}^{22}} & \overline{ U_{22}^{\,12}} & \overline{ U_{22}^{22}}
\end{array}
\right) $$

If $U$ is unitary then $U\, U^*=I$.
This relation implies  the following set of equations:

$$ 1)\, U_{11}^{11} \,\overline{ U_{11}^{11}} +  U_{11}^{\,1\,2} \, \overline{U_{11}^{\,1\,2}} +   U_{\,1\,2}^{11} \,\overline{ U_{\,1\,2}^{11}} +  U_{\,1\,2}^{\,1\,2} \, \overline{U_{\,1\,2}^{\,1\,2}}=1,$$

$$2)\, U_{11}^{11} \, \overline{U_{11}^{\,2\,1}} +  U_{11}^{\,1\,2} \,\overline{ U_{11}^{22} }+   U_{\,1\,2}^{11} \, \overline{U_{\,1\,2}^{\,2\,1}} +  U_{\,1\,2}^{\,1\,2} \,\overline{ U_{\,1\,2}^{22}}=0,$$

$$3)\, U_{11}^{11} \, \overline{U_{\,2\,1}^{11}} +  U_{11}^{\,1\,2} \, \overline{U_{\,2\,1}^{\,1\,2}} +   U_{\,1\,2}^{11} \,\overline{ U_{22}^{11} }+  U_{\,1\,2}^{\,1\,2} \, \overline{U_{22}^{\,1\,2}}=0,$$

$$ 4)\,U_{11}^{11} \, \overline{U_{\,2\,1}^{\,2\,1}} +  U_{11}^{\,1\,2} \,\overline{ U_{\,2\,1}^{22}} +   U_{\,1\,2}^{11} \, \overline{U_{22}^{\,2\,1}} +  U_{\,1\,2}^{\,1\,2} \,\overline{ U_{22}^{22}}=0,$$

$$ 5)\,U_{11}^{\,2\,1} \, \overline{U_{11}^{11}} +  U_{11}^{22} \,\overline{ U_{11}^{\,1\,2}} +   U_{\,1\,2}^{\,2\,1} \, \overline{U_{\,1\,2}^{11}} +  U_{\,1\,2}^{\,2\,2} \,\overline{ U_{\,1\,2}^{\,1\,2}}=0,$$

$$ 6)\,U_{11}^{\,2\,1} \, \overline{U_{11}^{\,2\,1}} +  U_{11}^{22} \,\overline{ U_{11}^{22}} +   U_{\,1\,2}^{\,2\,1} \, \overline{U_{\,1\,2}^{\,2\,1}} +  U_{\,1\,2}^{22} \,\overline{ U_{\,1\,2}^{22}}=1,$$

$$ 7)\,U_{11}^{\,2\,1} \,\overline{ U_{\,2\,1}^{11}} +  U_{11}^{22} \, \overline{U_{\,2\,1}^{\,1\,2}} +   U_{\,1\,2}^{\,2\,1} \,\overline{ U_{22}^{11} }+  U_{\,1\,2}^{22} \, \overline{U_{22}^{\,1\,2}}=0,$$

$$8)\,  U_{11}^{\,2\,1} \, \overline{U_{\,2\,1}^{\,2\,1}} +  U_{11}^{22} \,\overline{ U_{\,2\,1}^{22}}+   U_{\,1\,2}^{\,2\,1} \, \overline{U_{22}^{\,2\,1}} +  U_{\,1\,2}^{22} \, \overline{U_{22}^{22}}=0,$$

$$ 9)\,U_{\,2\,1}^{11} \, \overline{U_{11}^{11}} +  U_{\,2\,1}^{\,1\,2} \, \overline{U_{11}^{\,1\,2}} +   U_{22}^{11} \, \overline{U_{\,1\,2}^{11}} +  U_{22}^{\,1\,2} \, \overline{U_{\,1\,2}^{\,1\,2}}=0,$$

$$ 10)\,U_{\,2\,1}^{11} \, \overline{U_{11}^{\,2\,1}} +  U_{\,2\,1}^{\,1\,2} \, \overline{U_{11}^{22}} +   U_{22}^{11} \, \overline{U_{\,1\,2}^{\,2\,1}} +  U_{22}^{\,1\,2} \, \overline{U_{\,1\,2}^{22}}=0,$$

$$ 11)\,U_{\,2\,1}^{11} \, \overline{U_{\,2\,1}^{11}} +  U_{\,2\,1}^{\,1\,2} \, \overline{U_{\,2\,1}^{\,1\,2}} +   U_{22}^{11} \, U_{22}^{11} +  U_{22}^{\,1\,2} \, \overline{U_{22}^{\,1\,2}}=1,$$

$$12)\,  U_{\,2\,1}^{11} \, \overline{U_{\,2\,1}^{\,2\,1}} +  U_{\,2\,1}^{\,1\,2} \,\overline{ U_{\,2\,1}^{22}} +   U_{22}^{11} \, \overline{U_{22}^{\,2\,1}} +  U_{22}^{\,1\,2} \,\overline{ U_{22}^{22}}=0,$$

$$13)\,  U_{\,2\,1}^{\,2\,1} \, \overline{U_{\,2\,1}^{11}} +  U_{\,2\,1}^{22} \,\overline{ U_{\,2\,1}^{\,1\,2}} +   U_{22}^{\,2\,1} \, \overline{U_{22}^{11}} +  U_{22}^{22} \, \overline{U_{22}^{\,1\,2}}=0.$$

$$14)\,  U_{\,2\,1}^{\,2\,1} \, \overline{U_{11}^{11}} +  U_{\,2\,1}^{22} \,\overline{ U_{11}^{\,1\,2}} +   U_{22}^{\,2\,1} \, \overline{U_{\,1\,2}^{11}} +  U_{22}^{22} \, \overline{U_{\,1\,2}^{\,1\,2}}=0.$$

$$15)\,  U_{\,2\,1}^{\,2\,1} \, \overline{U_{11}^{\,2\,1}} +  U_{\,2\,1}^{22} \,\overline{ U_{11}^{22}} +   U_{22}^{\,2\,1} \, \overline{U_{\,1\,2}^{\,2\,1}} +  U_{22}^{22} \, \overline{U_{\,1\,2}^{22}}=0.$$

$$16)\,  U_{\,2\,1}^{\,2\,1} \, \overline{U_{\,2\,1}^{\,2\,1}} +  U_{\,2\,1}^{22} \,\overline{ U_{\,2\,1}^{22}} +   U_{22}^{\,2\,1} \, \overline{U_{22}^{\,2\,1}} +  U_{22}^{22} \, \overline{U_{22}^{22}}=1.$$

\medskip

Equation  2) is  equivalent to 5), equation 12) equivalent to 13), equation 8) equivalent to 15), equation 3) equivalent to 9), equation 7) equivalent to 10)  and equation 4) equivalent to 14). Then  we have $6$ free parameters for the coefficients of $U$.


Using the entries $U^{ij}_{rs}$ we considered above we define
$$\tilde{L}(Q)= p_1 \,\sum_{i=1}^2\, \left(\begin{array}{cc}
\overline{U}^{i1}_{11} & \overline{U}^{i1}_{\,21}\\
\overline{U}^{i1}_{\,12} & \overline{U}^{i1}_{22}
\end{array}
\right)
\,Q\,
\left(\begin{array}{cc}
U^{i1}_{11} & U^{i1}_{\,12}\\
U^{i1}_{\,21} & U^{i1}_{22}
\end{array}
\right)\,  +  $$
$$p_2 \,\sum_{i=1}^2\,\left(\begin{array}{cc}
\overline{U^{i2}_{11}} & \overline{U^{i2}_{\,21}}\\
\overline{U^{i2}_{\,12}} & \overline{U^{i2}_{22}}
\end{array}
\right)
\,Q\,
\left(\begin{array}{cc}
U^{i2}_{11} & U^{i2}_{\,12}\\
U^{i2}_{\,21} & U^{i2}_{22}
\end{array}
\right)\,
$$

We can consider an auxiliary $L_{ij}$ and express
$$ \tilde{L}(Q)= \sum_{i=1}^2 (\sqrt{p_1}\, (U^{i1})^*) \, Q \,  (\sqrt{p_1}\, U^{i1})+ \,\sum_{i=1}^2 (\sqrt{p_2}\, (U^{i2})^*) \, Q \,  (\sqrt{p_2}\, U^{i2})=$$
$$ \sum_{i=1}^2 L_{i1}^* \,Q \,L_{i1}  + \sum_{i=1}^2 L_{i2}^* \,Q \,L_{i2}=\sum_{i.j=1}^2 L_{ij}^* \,Q\, L_{ij}.$$

From the fact that $U\, U^* =I$ it follows (after a long computation) that
$$ \tilde{L}(I) =I.$$

Note that $ \tilde{L}$ preserve the cone of positive matrices.


Using the entries $U^{ij}_{rs}$ described above we
denote
$$ \hat{L}(Q)= \,p_1\,\sum_{i=1}^2\,\left(
\begin{array}{cc}
U^{i1}_{11} & U^{i1}_{\,12}\\
U^{i1}_{\,21} & U^{i1}_{22}
\end{array}
\right) \, Q \, \left(
\begin{array}{cc}
\overline{U^{i1}_{11}} & \overline{U^{i1}_{\,21}}\\
 \overline{U^{i1}_{\,12}} & \overline{U^{i1}_{22}}
\end{array}
\right)   +$$
$$ \,p_2\,\sum_{i=1}^2\,\left(
\begin{array}{cc}
U^{i2}_{11} & U^{i2}_{\,12}\\
U^{i2}_{\,21} & U^{i2}_{22}
\end{array}
\right) \, Q \, \left(
\begin{array}{cc}
\overline{U^{i2}_{12}} & \overline{U^{i2}_{\,21}}\\
 \overline{U^{i2}_{\,12}} & \overline{U^{i2}_{22}}
\end{array}
\right)=\sum_{i.j=1}^2 L_{ij} Q L_{ij}^* .  $$

\medskip
One can also show that $\hat{L} (Q)\,=\,Tr_2 [\,U\, (Q \otimes \beta)\, U^*\,] $ (see \cite{NP}).
\medskip

The first expression is the Kraus decomposition and the second the Stinespring dilation.

\medskip

Moreover $ \hat{L}$ preserve density matrices. This is proved in the appendix but we can present here another way to get that. If $Q$ is a density matrix, then
$$ Tr (\hat{L}(Q))= Tr  (\sum_{i.j=1}^2 L_{ij} Q L_{ij}^*) = \sum_{i.j=1}^2 Tr (L_{ij} Q L_{ij}^*) =\sum_{i.j=1}^2 Tr  (Q \, L_{ij}^* L_{ij} )=
 $$
 $$Tr  (\sum_{i.j=1}^2 Q \, L_{ij}^* L_{ij} )=Tr ( Q \,\sum_{i.j=1}^2 L_{ij}^* L_{ij})= Tr (Q)=1$$

 \bigskip

We denote
$$ Q =  \left(
\begin{array}{cc}
Q_{11} & Q_{12}\\
Q_{21} & Q_{22}
\end{array}
\right)\,.$$

Then,
$$U^{ij} \,\,Q \,\, (U^{ij})^*=\,\left(
\begin{array}{cc}
U^{ij}_{11} & U^{ij}_{\,12}\\
U^{ij}_{\,21} & U^{ij}_{22}
\end{array}
\right)  \left(
\begin{array}{cc}
Q_{11} & Q_{12}\\
Q_{21} & Q_{22}
\end{array}
\right)\, \left(
\begin{array}{cc}
\overline{U^{ij}_{11}} & \overline{U^{ij}_{\,21}}\\
 \overline{U^{ij}_{\,12}} & \overline{U^{ij}_{22}}
\end{array}
\right)   =$$





$$\left(
\begin{array}{cc}
  \overline{U^{ij}_{11}} ( U^{ij}_{11} Q_{11} +   U^{ij}_{\,12} Q_{21} )\,+\, \overline{U^{ij}_{12}} ( U^{ij}_{11} Q_{12} +   U^{ij}_{12} Q_{22} ) &\overline{U^{ij}_{\,21}} ( U^{ij}_{11} Q_{11} +   U^{ij}_{12} Q_{21} )\,
\,+\,\overline{ U^{ij}_{22} }( U^{ij}_{11} Q_{12} +   U^{ij}_{12} Q_{22} )    \\
\overline{U^{ij}_{11}} ( U^{ij}_{\,21} Q_{11} +   U^{ij}_{22} Q_{21} )\,+\, \overline{U^{ij}_{\,12}} ( U^{ij}_{\,21}\, Q_{12} +   U^{ij}_{22} Q_{22} ) &  \overline{U^{ij}_{21}} ( U^{ij}_{21} Q_{11} +   U^{ij}_{22} Q_{21} )\,+\, \overline{U^{ij}_{22}} ( U^{ij}_{21} Q_{12} +   U^{ij}_{22} Q_{22} )
\end{array}
\right),$$

We have to compute
$$ \hat{L} (Q) = p_1 \,[\, U^{11} \, Q \, (U^{11})^* + U^{21} \, Q\, (U^{21})^* \,] \,\,+ p_2 \,[\,U^{12} \, Q \, (U^{12})^* + U^{22} \, Q\, (U^{22})^* \,].$$

The coordinate $a_{11}$ of $ \hat{L} (Q) $ is
$$ p_1\,[\, \overline{U^{11}_{11}} ( U^{11}_{11} Q_{11} +   U^{11}_{\,12} Q_{21} )\,+\, \overline{U^{11}_{12}} ( U^{11}_{11} Q_{12} +   U^{11}_{12} Q_{22} ) \,]+$$
$$ p_1\,[\, \overline{U^{21}_{11}} ( U^{21}_{11} Q_{11} +   U^{21}_{\,12} Q_{21} )\,+\, \overline{U^{21}_{12}} ( U^{21}_{11} Q_{12} +   U^{21}_{12} Q_{22} ) \,]+$$
$$ p_2\,[\, \overline{U^{12}_{11}} ( U^{12}_{11} Q_{11} +   U^{12}_{\,12} Q_{21} )\,+\, \overline{U^{12}_{12}} ( U^{12}_{11} Q_{12} +   U^{12}_{12} Q_{22} ) \,]+$$
\begin{equation} \label{C} p_2\,[\, \overline{U^{22}_{11}} ( U^{22}_{11} Q_{11} +   U^{22}_{\,12} Q_{21} )\,+\, \overline{U^{22}_{12}} ( U^{22}_{11} Q_{12} +   U^{22}_{12} Q_{22} ) \,].\end{equation}

\bigskip

The coordinate $a_{12}$ is
$$ p_1\,[\, \overline{U^{11}_{21}} ( U^{11}_{11} Q_{11} +   U^{11}_{\,12} Q_{21} )\,+\, \overline{U^{11}_{22}} ( U^{11}_{11} Q_{12} +   U^{11}_{12} Q_{22} ) \,]+$$
$$ p_1\,[\, \overline{U^{21}_{21}} ( U^{21}_{11} Q_{11} +   U^{21}_{\,12} Q_{21} )\,+\, \overline{U^{21}_{22}} ( U^{21}_{11} Q_{12} +   U^{21}_{12} Q_{22} ) \,]+$$
$$ p_2\,[\, \overline{U^{12}_{21}} ( U^{12}_{11} Q_{11} +   U^{12}_{\,12} Q_{21} )\,+\, \overline{U^{12}_{22}} ( U^{12}_{11} Q_{12} +   U^{12}_{12} Q_{22} ) \,]+$$
\begin{equation} \label{C} p_2\,[\, \overline{U^{22}_{21}} ( U^{22}_{11} Q_{11} +   U^{22}_{\,12} Q_{21} )\,+\, \overline{U^{22}_{22}} ( U^{22}_{11} Q_{12} +   U^{22}_{12} Q_{22} ) \,].\end{equation}

We will consider a parametrization of the density matrices taking $Q_{11}= 1-Q_{22}$ and $Q_{12} = \overline{Q_{21}}$.

The variable $Q_{11}$ is positive in the real line and smaller than one. Indeed, by positivity of $Q$, we have $0\leq Q_{11} Q_{22}= Q_{11} (1- Q_{11})= Q_{11} - Q_{11}^2.$

$Q_{12} $ is in $\mathbb{C}= \mathbb{R}^2$ but satisfying $Q_{11} (1- Q_{11}) - Q_{12} \overline{Q}_{12} \geq  0 $ because we are interested in density matrices which are positive operators.

The numbers $p_1$ and $p_2$ are fixed. Consider the function $G$ such that
$$G(Q_{11},Q_{12}) \,=\, $$
$$ (\,\, p_1\,[\, \overline{U^{11}_{11}} ( U^{11}_{11} Q_{11} +   U^{11}_{\,12} \overline{Q_{12}} )\,+\, \overline{U^{11}_{12}} ( U^{11}_{11} Q_{12} +   U^{11}_{12} (1- Q_{11}) ) \,]+$$
$$ p_1\,[\, \overline{U^{21}_{11}} ( U^{21}_{11} Q_{11} +   U^{21}_{\,12} \overline{Q_{12}} )\,+\, \overline{U^{21}_{12}} ( U^{21}_{11} Q_{12} +   U^{21}_{12} (1-Q_{11}) ) \,]+$$
$$ p_2\,[\, \overline{U^{12}_{11}} ( U^{12}_{11} Q_{11} +   U^{12}_{\,12} \overline{Q_{12}} )\,+\, \overline{U^{12}_{12}} ( U^{12}_{11} Q_{12} +   U^{12}_{12} (1- Q_{11}) ) \,]+$$
$$ p_2\,[\, \overline{U^{22}_{11}} ( U^{22}_{11} Q_{11} +   U^{22}_{\,12} \overline{ Q_{12}} )\,+\, \overline{U^{22}_{12}} ( U^{22}_{11} Q_{12} +   U^{22}_{12} (1- Q_{11}) ) \,]\,\,\,\,\,,$$
$$ \,\, p_1\,[\, \overline{U_{21}^{11}} ( U^{11}_{11} Q_{11} +   U^{11}_{\,12} \overline{Q_{12}} )\,+\, \overline{U^{11}_{22}} ( U^{11}_{11} Q_{12} +   U^{11}_{12} (1- Q_{11}) ) \,]+$$
$$ p_1\,[\, \overline{U^{21}_{21}} ( U^{21}_{11} Q_{11} +   U^{21}_{\,12} \overline{Q_{12}} )\,+\, \overline{U^{21}_{22}} ( U^{21}_{11} Q_{12} +   U^{21}_{12} (1-Q_{11}) ) \,]+$$
$$ p_2\,[\, \overline{U^{12}_{21}} ( U^{12}_{11} Q_{11} +   U^{12}_{\,12} \overline{Q_{12}} )\,+\, \overline{U^{12}_{22}} ( U^{12}_{11} Q_{12} +   U^{12}_{12} (1- Q_{11}) ) \,]+$$
$$ p_2\,[\, \overline{U^{22}_{21}} ( U^{22}_{11} Q_{11} +   U^{22}_{\,12} \overline{ Q_{12}} )\,+\, \overline{U^{22}_{22}} ( U^{22}_{11} Q_{12} +   U^{22}_{12} (1- Q_{11}) ) \,]\,\,)$$

\medskip

When there is  a  unique fixed point for $G$?
\bigskip

{\bf Example:}
Suppose $U= e^{i\,\beta\,\sigma^x \otimes \sigma^x}= $$\cos(\beta) \, (I \otimes I)\,+\,i\, \sin(\beta)\, (\sigma_x \otimes \sigma_x).$ In this case
$$U=   \left(
\begin{array}{cccc}
\cos \beta & 0 &  0 & i\, \sin \beta\\
 0 &\cos \beta   &  i\, \sin \beta & 0\\
0 & i\, \sin \beta &   \cos \beta  & 0\\
i\, \sin \beta & 0 &  0 & \cos \beta
\end{array}
\right) $$

Therefore,

$$G(Q_{11},Q_{12}) \,=\, $$
$$ (\,(\, p_1 - p_1 Q_{11}  \,+ p_2\,- p_2 Q_{11})  \,,$$
$$ \,\, p_1\,\, (\cos \beta)^2  Q_{12}  \,+p_1\,\, (\sin \beta)^2 \overline{Q_{12}} \, \,+ p_2\,\,(\sin \beta)^2 \, \overline{Q_{12}}\,+ p_2\,\, (\cos \beta)^2 \, Q_{12} \,\,)=$$
$$ (\,\,1-Q_{11}   \,,\,\, p_1\,\, (\cos \beta)^2  Q_{12}  \,+p_1\,\, (\sin \beta)^2 \overline{Q_{12}} \, \,+ p_2\,\,(\sin \beta)^2 \, \overline{Q_{12}}\,+ p_2\,\, (\cos \beta)^2 \, Q_{12} \,\,)$$

One can easily see that given any $a\in \mathbb{R}$ we have that $Q_{11} =1/2$, and $Q_{12}=a$ determine  a fixed point for $G$.
In order the fixed point matrix  to be positive we need that $-1/2< a<1/2$.

In this case the fixed point is not unique.

\bigskip

It is more convenient to express $G$ in terms of
the variables $Q_{11}\in [0,1]$, and $(a,b) \in \mathbb{R}^2$, where $Q_{12}= a + b i$. As these parameters describes density matrices there are some restrictions: $1/4\,\geq Q_{11} (1- Q_{11}) \geq (a^2 + b^2)$  and $1\geq Q_{11} \geq 0$

We denote by $Re (z)$ the real part of the complex number $z$ and by $Im(z)$ its imaginary part.

In this case we get
$$ G(Q_{11},a,b)=$$
$$( Q_{11} \alpha_1 + \beta_1\,+ (a_{11} + a_{12})\, a +\,i \,  (a_{11} -a _{12}) \,b \,,\,$$
$$Re \,(\,Q_{11} \alpha_2 + \beta_2 \,+\,(a_{21} + a_{22})\, a +\,i \,  (a_{21} -a _{22}) \,b\,)\,, $$
$$Im \,(\,Q_{11} \alpha_2 + \beta_2 \,+\,(a_{21} + a_{22})\, a +\,i \,  (a_{21} -a _{22}) \,b\,)\,). $$

where
$$\alpha_1= p_1 [\,\overline{ U^{11}_{ 11}}  U^{11}_{ 11}- \overline{  U^{11}_{ 12}}  U^{11}_{ 12}  + \overline{  U^{21}_{ 11 }} U^{21}_{ 11} -
\overline{ U^{21}_{ 12 }} U^{21}_{ 12}\,]\,+\, $$
$$  p_2 [\,\overline{ U^{12}_{ 11}}  U^{12}_{ 11}- \overline{  U^{12}_{ 12}}  U^{12}_{ 12}  + \overline{  U^{22}_{ 11 }} U^{22}_{ 11} -
\overline{ U^{22}_{ 12 }} U^{22}_{ 12}\,]\,\, ,$$

,
$$\beta_1= p_1 [\,\overline{ U^{11}_{ 12}}  U^{11}_{ 12}+ \overline{  U^{21}_{ 12}}  U^{21}_{ 12} \,]  + p_2\,[\,\overline{  U^{12}_{ 12 }} U^{12}_{ 12} +\overline{ U^{22}_{ 12 }} U^{22}_{ 12}\,]\,\, ,$$

$$\alpha_2= p_1 [\,\overline{ U_{21}^{ 11}}  U^{11}_{ 11}- \overline{  U^{11}_{ 22}}  U^{11}_{ 12}  + \overline{  U^{21}_{ 21 }} U^{21}_{ 11} -
\overline{ U^{21}_{ 22 }} U^{21}_{ 12}\,]\,+\, $$
$$  p_2 [\,\overline{ U^{12}_{ 21}}  U^{12}_{ 11}- \overline{  U^{12}_{ 22}}  U^{12}_{ 12}  + \overline{  U^{22}_{ 21 }} U^{22}_{ 11} -
\overline{ U^{22}_{ 22 }} U^{22}_{ 12}\,]\,\, ,$$

$$\beta_2= p_1 [\,\overline{ U^{11}_{ 22}}  U^{11}_{ 12}+ \overline{  U^{21}_{ 21}}  U^{21}_{ 12} \,]  + p_2\,[\,\overline{  U^{12}_{ 22 }} U^{12}_{ 12} +\overline{ U^{22}_{ 22 }} U^{22}_{ 12}\,]\,\, ,$$

$$a_{11} = p_1 [\,\overline{ U^{11}_{ 12}}  U^{11}_{ 11}+ \overline{  U^{21}_{ 12}}  U^{21}_{ 11} \,]  + p_2\,[\,\overline{  U^{12}_{ 12 }} U^{12}_{ 11} +\overline{ U^{22}_{ 12 }} U^{22}_{ 11}\,]\,\, ,$$

$$a_{12} = p_1 [\,\overline{ U^{11}_{ 11}}  U^{11}_{ 12}+ \overline{  U^{21}_{ 11}}  U^{21}_{ 12} \,]  + p_2\,[\,\overline{  U^{12}_{ 11 }} U^{12}_{ 12} +\overline{ U^{22}_{ 11 }} U^{22}_{ 12}\,]\,\, ,$$

$$a_{21} = p_1 [\,\overline{ U^{11}_{22}}  U^{11}_{ 11}+ \overline{  U^{21}_{ 22}}  U^{21}_{ 11} \,]  + p_2\,[\,\overline{  U^{12}_{ 22 }} U^{12}_{ 11} +\overline{ U^{22}_{ 22 }} U^{22}_{ 11}\,]\,\, ,$$

$$a_{22} = p_1 [\,\overline{ U_{21}^{11}}  U^{11}_{ 12}+ \overline{  U^{21}_{ 21}}  U^{21}_{ 12} \,]  + p_2\,[\,\overline{  U^{12}_{ 21 }} U^{12}_{ 12} +\overline{ U^{22}_{ 21 }} U^
{22}_{ 12}\,]\,\, ,$$

$\alpha_1$ is  a real number. As $\Phi$ takes density matrices to density matrices we have that $\beta_1$ is also real.

Note that $|\alpha_1|<1$ and $1>\beta_1>0.$

It is easy to see from the above equations that  $(a_{11} + a_{12})$ and $\,i \,  (a_{11} -a _{12}) $ are both real numbers.

We are not able to say the same for $(a_{21} + a_{22})\, a $ or $\,i \,  (a_{21} -a _{22}) \,b\,\,.$

In order to find the fixed point we have to solve
$$ Q_{11} \alpha_1 + \beta_1\,+ (a_{11} + a_{12})\, a +\,i \,  (a_{11} -a _{12}) \,b \,= Q_{11}\,$$
$$ Q_{11} \alpha_2 + \beta_2 \,+\,(a_{21} + a_{22})\, a +\,i \,  (a_{21} -a _{22}) \,b\,\,= a + bi ,$$

which means in matrix form

$$ \left(
\begin{array}{ccc}
(\alpha_1-1)  & a_{11} + a_{12} & i\,(a_{11} - a_{12})\\
 \alpha_2 & a_{21} + a_{22} - 1   &  i\, (a_{21} - a_{22} -\,1)
\end{array}
\right)    \left(
\begin{array}{c}
Q_{11} \\
a\\
b
\end{array}
\right)=   \left(
\begin{array}{c}
- \beta_1 \\
- \beta_2\end{array}
\right).$$

We are interested in real solutions $Q_{11},a,b$.

 In the case of the example mentioned above one can show that $\alpha_1=1$ and $\alpha_0=0$ which means that in the expressions above
 we get a set of two equation in  two variables $a,b$,

 Remember that we are interested in matrices such that $1/4\,\geq Q_{11} (1- Q_{11}) \geq (a^2 + b^2).$
Notice that $0\leq Q_{11} \leq 1$. As $\Phi$ takes density matrices to density matrices there is a fixed point for $G$ by the Brower fixed point theorem. The main question is the conditions on $U$ and $\beta$ such that the fixed point is unique.

If there is a solution $(\hat{Q}_{11}, \hat{a}, \hat{b})\neq (0,0,0)$ in $\mathbb{R}^3$ to the equations

$$ \hat{Q}_{11} (\alpha_1 - 1) + (a_{11} + a_{12})\, \hat{a} +\,i \,  (a_{11} -a _{12}) \,\hat{b} \,= 0\,$$
\begin{equation} \label{es}\,\hat{Q}_{11} \alpha_2  \,+\,(a_{21} + a_{22}- 1)\, \hat{a} +\,i \,  (a_{21} -a _{22}-1) \,\hat{b}\,\,=  0,\end{equation}
then, the fixed point is not unique. The condition is necessary and sufficient.

A necessary condition for the fixed point to be unique is to be nonull the determinant of the operator
$$\,\, K= \left(
\begin{array}{ccc} a_{11} + a_{12} & i\,(a_{11} - a_{12})\\
 a_{21} + a_{22} - 1   &  i\, (a_{21} - a_{22} -\,1)
\end{array}
\right)    .$$

Notice that if $(z_1,z_2)$ satisfies $K(z_1,z_2)=(0,0)$, then $\frac{z_1}{z_2}$ is real (because $a_{11} + a_{12} $ and $i\,(a_{11} - a_{12})$ are real). From this follows that there  exist a  solution $(a,b)\in \mathbb{R}^2$ in the kernel of $K$. In this case $(0,a,b)$ is a nontrivial solution
of (\ref{es}).

The condition det $K\neq 0$ is an open and dense property on the  unitary matrices $U$. Indeed,  there are six free parameters on the coefficients $U_{rs}^{ij}$. Consider an initial unitary operator $U$. One can fix 5 of them  and move a little bit the last one. This will change $U$ and will move the determinant of $K_U$ in such way that can avoid the value $0$ for some small perturbation of the initial $U$.

Suppose $U$ satisfies such property Det $U\neq 0$.  For each real value $Q_{11}$ we get a different $(a_{Q_{11}},b_{Q_{11}})$ which is a solution
of $K(a,b)= (-\,Q_{11} (\alpha_1-1), -\,Q_{11}\, \alpha_2)$.

In this way we get an infinite number of solutions  $(Q_{11},a_{Q_{11}},b_{Q_{11}})\in \mathbb{R} \times \mathbb{C}^2$ to (\ref{es}).

$\alpha_2$ is not real.

But, we need solutions on  $\mathbb{R}^3$. Denote by $S=S_U$ the linear subspace of vectors in $\mathbb{C}^2$ of the form
$\rho (\alpha_1 -1 , \alpha_2),$ where $\rho$ is complex.

\begin{lemma} For an open and dense set of unitary $U$ we  get that $K^{-1} (S)\, \cap\, \mathbb{ R}^2= \{(0,0)\}.$ For such $U$, suppose $(Q_{11},a,b) $ satisfies equation (\ref{es}), then the non-trivial  solutions $(\hat{a},\hat{b})$ of
$$K(\hat{a},\hat{b})= (-\,Q_{11} (\alpha_1-1), -\,Q_{11}\, \alpha_2)$$
are not in $\mathbb{R}^2$.

\end{lemma}
{\bf Proof:} Suppose $\frac{ 1-\alpha_1}{\alpha_2} = \alpha + \beta i= z^0=z_U^0.$ Note that for a generic $U$ we have that $\alpha_2\neq 0.$

We denote $C_{11} = a_{11} + a_{12}$, $C_{12} = i\,(a_{11} - a_{12})$, $C_{21} = a_{21} + a_{22}-1 $ and finally $C_{22} =i\, ( a_{21} - a_{22}- 1 ) $.

Suppose $(Q_{11},a,b) \in \mathbb{R}^3$ satisfies equation (\ref{es}). We know that generically on $U$ the value $Q_{11}$ is not zero.

For each $C_{ij}$ we denote $C_{ij} = C_{ij}^1 + C_{ij}^2 \,\,i$, where $\,i,j=1,2$.

If $K(\hat{a},\hat{b})= (-\,Q_{11} (\alpha_1-1), -\,Q_{11}\, \alpha_2)$, then
$$ C_{11} \hat{a} + C_{21} \hat{b}= z^0\,( C_{21} \hat{a} + C_{22} \hat{b})= (\alpha + \beta i)\,  ( C_{21} \hat{a} + C_{22} \hat{b}).$$

In this case
$$ C_{11} \hat{a} + C_{21} \hat{b}= (\alpha C_{21}^1 \hat{a}  - \beta C_{21}^2 \hat{a} - \beta C_{22}^1 \hat{b}- \alpha C_{22}^2 \hat{b}) + $$
$$ i\, (\beta C_{21}^1 \hat{a}  + \alpha C_{21}^2 \hat{a} + \alpha C_{22}^1 \hat{b}- \beta C_{22}^2 \hat{b}).$$

If $\hat{a}$ and $\hat{b}$ are real, then, as $C_{11}$ and $C_{22}$ are real , then
\begin{equation} \label{v1} (\beta C_{21}^1   + \alpha C_{21}^2)\, \hat{a} + (\alpha C_{22}^1 - \beta C_{22}^2) \hat{b}=0.\end{equation}

Moreover,

\begin{equation} \label{v2}(\alpha C_{21}^1   - \beta C_{21}^2- C_{11})\, \hat{a} - (\beta C_{22}^1 - \alpha C_{22}^2  - C_{21} )\,\hat{b}=0 \end{equation}

If

$$\,\,\text{Det}\,\,\left(
\begin{array}{ccc} \beta C_{21}^1   + \alpha C_{21}^2& \alpha C_{22}^1 - \beta C_{22}^2\\
 \alpha C_{21}^1   - \beta C_{21}^2- C_{11}  & \beta C_{22}^1 - \alpha C_{22}^2  - C_{21}
\end{array}
\right)   \neq 0 ,$$

then just the trivial solution $(0,0)$ satisfies (\ref{v1}) and (\ref{v2}).

The above determinant is non zero in an open and dense set of $U$.

Then, the solution $(Q_{11},a,b) \in \mathbb{R}^3$ of (\ref{es}) have to be trivial.
\qed

\bigskip

Under this two assumptions on $U$ (which are open and dense) the fixed point for $G$ is unique.
Then, it follows that the density matrix $Q=Q_\Phi$ which is invariant for $\Phi$ is unique.
Given an initial $Q_0$ any convergent subsequence $\Phi^{n_k}(Q_0),$ $\k \to \infty$ will converge to the fixed point (because is unique).

As
$$ G(Q_{11},a,b)=$$
$$( Q_{11} \alpha_1 + \beta_1\,+ (a_{11} + a_{12})\, a +\,i \,  (a_{11} -a _{12}) \,b \,,\,$$
$$Re \,(\,Q_{11} \alpha_2 + \beta_2 \,+\,(a_{21} + a_{22})\, a +\,i \,  (a_{21} -a _{22}) \,b\,)\,, $$
$$Im \,(\,Q_{11} \alpha_2 + \beta_2 \,+\,(a_{21} + a_{22})\, a +\,i \,  (a_{21} -a _{22}) \,b\,)\,), $$
one can find the explicit solution
$$Q_\Phi=  \left(
\begin{array}{cc}
Q_{11} & a+ b \, i\\
a-b\, i& 1- Q_{11}
\end{array}
\right)\,$$

by solving the  linear problem $G(Q_{11},a,b)=(Q_{11},a,b)$.


\section{Appendix}

\begin{lemma} \label{tet} Given $A,B\in \mathcal{L}(V)$, then $Tr (A \otimes B) = Tr (\, Tr_2 (A \otimes B)\,).$ Moreover, $Tr(\, Tr_2 (T)\,)= Tr(T)$, for all $T \in  \mathcal{L}(V\otimes V).$

\end{lemma}

{\bf Proof:}

Indeed,
$$ Tr (A \otimes B) = Tr (A)\, Tr (B) = Tr (\, Tr(A)\, B)= Tr (\,Tr_2 (A \otimes B)\,).$$

\qed

\begin{lemma} \label{eset} Given $T \in  \mathcal{L}(V\otimes V),$

a) if $T$ is selfadjoint, then, $Tr_2$ is also selfadjoint,

b) moreover, if $T$ is also positive semidefinite then $Tr_2(T)$ is  semidefinite.

\end{lemma}

{\bf Proof:}

a) If $T$ is selfadjoint, then, $t_{ijkl}= \overline{t_{klij}}$. This implies that
$\sum_j t_{ijkj}=\sum_j  \overline{t_{kjij}}.$ Therefore, $Tr_2$ is  selfadjoint.
\medskip

b) If $T$ is postive semidefinite, then $< \,T(x \otimes x')\,,\, x \otimes x'\,>\,\geq 0,$ for all $x,x'\,\in V$. In particular,
$< \,T(x \otimes e_q)\,,\, x \otimes e_q\,>\,\geq 0,$ for all $x= c_1\, e_1+....+c_n e_n\,\in V$ and $1 \leq q \leq n.$

As $T(x\otimes e_q) = \sum t_{ijkl}\, L_{ik}(x) \otimes L_{jl} (e_q)= \sum t_{ijkq}\,c_k \,(e_i \otimes e_j)$., then
$$ < \,T(x \otimes e_q)\,,\, x \otimes e_q\,>\,= \sum_{i,k} t_{iqkq}\,c_k \,\overline{c_i},\,\,\,\,\,\,q=1,2,...,n.$$

From this follows that $ \sum_{i,k,q} t_{iqkq}\,c_k \,\overline{c_i}= \sum_{i,k}\,\,\,(\,\sum_q t_{iqkq}\,)\,\, \,\,c_k\,\overline{c_i}\geq 0.$

Then, $< Tr_2 (T)\,(x),x>\geq 0.$

 \qed

Note that the analogous property for positive definite $T$ is also true.

\begin{lemma} If $A \in \Gamma$, then $\Phi_U(A)\in \Gamma$, for all $U \in \mathcal{U}.$

\end{lemma}
{\bf Proof:}

As $A$ and $\beta$ are selfadjoint and positive semidefinite the same is true for $A \otimes \beta.$ Then, the same is true for $U (A \otimes \beta)U^*.$ From Lemma
\ref{eset} we get that $\Phi_U (A) = Tr_2 (\,U ( A \otimes \beta)U^*\,)$ is selfadjoint.

By Lemma \ref{tet} $Tr (\Phi_U (A)) = Tr  (\,U ( A \otimes \beta)U^*)= Tr(A \otimes b)= Tr(A)\, TR(B)=1.$

\qed

\bigskip

Instituto de Matematica-UFRGS 

Av. Bento Gon\c calves, 9500. CEP 91509-900, Porto Alegre, Brasil
\bigskip

A. O. Lopes partially supported by CNPq, PRONEX -- Sistemas
Din\^amicos, INCT, and beneficiary of CAPES financial support

\bigskip

email: arturoscar.lopes$@$gmail.com


\begin{thebibliography}{99}




\bibitem{Real}  J. Bochnak, M. Coste and M-F Roy, Real Algebraic Geometry, Springer Verlag (1998)

\bibitem{Liu} C. Liu,
Unitary conjugation channels with continuous random phases, Quantum Stud.: Math. Found. (2014)


\bibitem{NP} I. Nechita and C. Pellegrini, Random repeated quantum interactions and random invariant states, Probrab. Theory and Relat. Fields, 52 (2012) 299--320

\bibitem{BJK} L. Bruneau, A. Joye, and M. Merkli,
Repeated interactions in open quantum systems, Journal of Mathematical Physics 55, 075204 (2014)

\end{thebibliography}
\end{document}